\documentclass[12pt,preprint]{aastex}

\shorttitle{Solar Cycle Variation of Real CME Latitudes}
\shortauthors{Song W.B., Feng X.S., \& Hu Y.Q.}
\begin{document}

\title{Solar Cycle Variation of Real CME Latitudes}
\author{Wenbin Song, Xueshang Feng and Yanqi Hu}
\affil{State Key Laboratory for Space Weather, Center for Space
Science and Applied Research, Chinese Academy of Sciences, Beijing
100080, China} \email{wbsong@spaceweather.ac.cn}

\begin{abstract}

With the assumption of radial motion and uniform longitudinal
distribution of coronal mass ejections (CMEs), we propose a method
to eliminate projection effects from the apparent observed CME
latitude distribution. This method has been applied to SOHO LASCO
data from 1996 January to 2006 December. As a result, we find that
the real CME latitude distribution had the following
characteristics: (1) High-latitude CMEs ($\theta>60^{\circ}$ where
$\theta$ is the latitude) constituted 3\% of all CMEs and mainly
occurred during the time when the polar magnetic fields reversed
sign. The latitudinal drift of the high-latitude CMEs was
correlated with that of the heliospheric current sheet. (2) 4\% of
all CMEs occurred in the range
$45^{\circ}\leq\theta\leq60^{\circ}$. These mid-latitude CMEs
occurred primarily in 2000, near the middle of 2002 and in 2005,
respectively, forming a prominent three-peak structure; (3) The
highest occurrence probability of low-latitude ($\theta<
45^{\circ}$) CMEs was at the minimum and during the declining
phase of the solar cycle. However, the highest occurrence rate of
low-latitude CMEs was at the maximum and during the declining
phase of the solar cycle. The latitudinal evolution of
low-latitude CMEs did not follow the Sp\"{o}rer sunspot law, which
suggests that many CMEs originated outside of active regions.

\end{abstract}


\section{Introduction}

Coronal mass ejections (CMEs) are observed with white-light
coronagraphs as significant changes in coronal structure moving
outward (Skirgiello, 2003). They are believed to be the prime
driver of most space weather events, e.g., interplanetary shocks,
high-energy particles and geomagnetic storms (Gosling, 1993).
Recently many studies have been carried out on the properties of
CME source regions in order to gain insight into the initiation
mechanism of such phenomenon. Subramanian and Dere (2001) found
that 41\% of CME-related transients observed are associated with
active regions and have no filament eruptions, 44\% are associated
with eruptions of filaments embedded in active regions, and 15\%
are associated with eruptions of filaments outside active regions.
Zhou et al. (2003) found that 88\% of the earth-directed halo CMEs
are associated with flares and 94\% are associated with eruptive
filaments. With regard to the locations of CME source regions,
they found that there are 79\% CMEs coming from active regions and
21\% originating outside active regions. In this letter we
concentrate on the latitudinal variations of CME source regions
also with a view of providing clues about the CME triggering
processes. Skirgiello (2003) proposed a mathematical tool using
the apparent latitude distribution to deduce the real latitude
distribution. By referring to this idea, a much simpler and easier
new method with the same function is designed in order to obtain
the evolution of real CME latitudes during solar cycle 23.

\section{Data}

CMEs are selected from the catalog on the website of
\emph{http://cdaw.gsfc.nasa.gov/CME\_list} in the interval 1996
January to 2006 December, as observed by SOHO/LASCO. The
SOHO/LASCO coronagraph images solar corona continuously since 1996
covering a field of view starting from about 1.5 $R_s$ to 32 $R_s$
(Gopalswamy et al. 2003, $R_s$ is the sun's radius). By performing
preliminary estimation, St. Cyr et al. (2000) found that its
detection of CMEs possessed an efficiency not less than 95\%. The
CMEs are observed on the plane of the sky, therefore all apparent
spatial parameters are a projection of the real values onto that
plane. The apparent latitude $\delta$ comes from the central
position angle ($CPA$), which is defined and listed in the CME
catalog by Yashiro et al. When $0^{\circ}\leq CPA<180^{\circ}$,
$\delta=90^{\circ}-CPA$ and when $180^{\circ}\leq
CPA<360^{\circ}$, $\delta=CPA-270^{\circ}$. The complete halo
events with an angular width of $360^{\circ}$ are excluded from
the statistics because of difficulty in determining $\delta$.

In consideration of the CME number and the temporal precision, we
divide the whole interval into 22 parts, as shown in Figure 1. If
the time of a part is set very short, the CME number will get too
small to be able to represent the real distribution of $\delta$;
if very long, the temporal precision will then get so bad that it
is difficult to show the solar cycle variation clearly. For the
time interval of every part, we compute the occurrence percentages
of CMEs corresponding to $\delta=0^{\circ}$, $\delta=1^{\circ}$,
$\ldots$, and $\delta=90^{\circ}$, respectively. The bin width of
$\delta$ is $1^{\circ}$ by following that of the $CPA$. Hence we
can define a vector
\begin{equation}
S=[s_0,s_1,\ldots,s_i,\ldots,s_{90}]^T
  \label{1}
\end{equation}
to give the distribution of $\delta$, where $s_i$ means the
percentage at $\delta=i^{\circ}$. In the present study, $S$ is
smoothed with a window of $10^{\circ}$, see the thin solid lines
in Figure 1. Furthermore, the distributions for the northern
($\delta>0$) and southern ($\delta<0$) hemisphere are summed
without the consideration of the north-south asymmetry of solar
activities, e.g., the sunspots (Oliver \& Ballester, 1994), the
filaments (Hansen \& Hansen, 1975), the X-ray flares (Li et al.
1998) and the high-latitude CMEs (Gopalswamy et al. 2003).

\section{Method}

The proposed method is based on the assumption of radial motion
and uniform longitudinal distribution of CMEs. As shown in Figure
2, the point $E$ indicates the CME source region, $OE$ is the
direction of CME motion, $OH$ is $OE$'s projection onto the plane
of the sky. Thereupon $\angle{HOY}$, $\angle{LOY}$, and
$\angle{ACE}$ are, namely, the apparent latitude $\delta$, the
real latitude $\theta$ and the longitude $\varphi$. By a simple
deduction ($|CH|=|CE|\sin\varphi=|OC|/\tan\delta$,
$|CE|=R_s\cos\theta$, $|OC|=R_s\sin\theta$) we have
\begin{equation}
  \tan\theta=\sin\varphi\tan\delta.
  \label{1}
\end{equation}

Set the discrete variable
$\theta=[0^{\circ},1^{\circ},2^{\circ},\ldots,90^{\circ}]$. For each
$\theta$ fixed,  we can obtain 91 $\delta$ values by using Equation
2 if given $\varphi=0^{\circ}$, $\varphi=1^{\circ}, \cdots,
\varphi=90^{\circ}$.
 We let all $\delta$ equal to the expression
$\delta+0.5^{\circ}$ converted to integer type and then we can
define a vector $A=[a_0, a_1, \ldots, a_i, \ldots, a_{90}]^T$,
where $a_i$ means the percentage at $\delta=i^{\circ}$ in the
obtained 91 $\delta$ values. Some special conditions are dealt
with as follows. When $\theta=0^{\circ}$, $A=[1,0,\ldots,0]^T$;
when $\theta=90^{\circ}$, $A=[0,\ldots,0,1]^T$; when
$\varphi=0^{\circ}$ and $\theta\neq 0^{\circ}$,
$\delta=90^{\circ}$.

Combining 91 vectors $A$ we get a $91\times91$ matrix
\begin{equation}
M=[A_0,A_1,\ldots,A_j,\ldots,A_{90}],
  \label{1}
\end{equation}
where the subscript $j$ means the case $\theta=j^{\circ}$. Figure 3
shows the contour of matrix $M$. If we use vector
$R=[r_0,r_1,\ldots,r_{90}]^T$ to indicate the real distribution of
$\theta$, we have $M\times R=S$ and
\begin{equation}
R=M^{-1}\times S.
  \label{1}
\end{equation}

\section{Results and Discussion}

The final results of $R$, also smoothed with a window of
$10^{\circ}$, are plotted as 22 thick solid lines in Figure 1. In
comparison with $S$, most high-latitude CMEs are substituted with
events coming from low latitudes. The reason of this phenomenon is
that by projection the CMEs remote from solar limb can be observed
at any $\delta>\theta$. For a direct demonstration of the solar
cycle variation of $\theta$, we combine all vectors $R$ in time
order and give a filled contour plot in Figure 4a. Figure 4b then
shows the variation of the actual (not the percentage) CME
occurrence rate $\eta=R\times n/t$, where $n$ is the CME number,
$t$ is the time length (in $year$), both given in Figure 1.

From Figure 4, we find that most high-latitude CMEs occur during
the period from 1999 to the middle of 2001. Gopalswamy et al.
(2003) found a general spreading of CME latitudes attaining
$60^{\circ}$ by 1999. The northern high-latitude CMEs became
nonexist beyond October 2000. However the southern ones continued
to occur until the first quarter of 2002. Our time interval of
most high-latitude CMEs' occurrence is approximately consistent
with the above Gopalswamy et al.'s. As shown by white dotted
lines, we find a poleward motion with a speed of 12.5 $deg/year$
and an equatorward motion with a speed of 25.4 $deg/year$ taking
place during years 1999-2001 and during 2000 to the middle of
2001, respectively. For the poleward motion, years 1999 and 2001
are the polarity reversal time at $\theta=60^{\circ}$ and the
epoch when the polar regions' area is of minimum (Song \& Wang,
2006). Therefore we suggest this motion come from the collision
between the unipolar poleward meridional flows (Song \& Wang,
2006) and the polar regions of an opposite polarity. Such
collision hastens the formation of polar crown filaments whose
eruptions have a close correspondence with CMEs (Gopalswamy et al.
2003). The much lower speed than that of the meridional flows
($>20$ $deg/year$) can be explained by the polar regions blocking
the way. The equatorward motion may have the same mechanism
because during its interval, we can see that the tilt angle,
defined as the maximum extent of the heliospheric current sheet
(HCS), also declines sharply with a similar speed (see Figure 8 in
Gopalswamy et al. 2003). The high-latitude CMEs constitute 5-10\%
during the time when the polar magnetic fields reverse sign and
3\% of all CMEs during the whole solar cycle.

In the range $\theta\in[45^{\circ},60^{\circ}]$, we only find
three prominent peaks located in 2000, near the middle of 2002 and
in 2005, respectively, which seem to have a period of 2.5 $years$.
By comparing these peak positions with the variation of sunspot
number (see the pink solid lines in Figure 4), we  find no any
close correlations. Therefore this phenomenon is very puzzling
indeed. Wang et al. (1989) found many poleward surges of
alternating polarities in the active belts. Do they initiate the
middle-latitude CMEs? We have no definite answer at present. Such
CMEs constitute 6-12\% during their main phases and 4\% of all
CMEs.

93\% CMEs occur in the range $\theta<45^{\circ}$. From Figure 4,
we find that the highest occurrence probability ($R$) of these
low-latitude CMEs is at the minimum and during the declining phase
of solar cycle. However the highest occurrence rate ($\eta$) is at
the maximum and during the declining phase of solar cycle. As
shown by green solid lines, at the minimum and during the rising
phase, CMEs' average latitude has a linear increase. Then at the
maximum and during the declining phase, CMEs' average latitude is
in a nearly steady state, which keeps around
$15^{\circ}\sim25^{\circ}$. It is obvious that it does not follow
the Sp\"{o}rer sunspot law. This means that there are a large
quantity of CMEs originating outside active regions. Zhou et al.
(2006) classified CME-associated large-scale structures into four
different categories: extended bipolar regions (EBRs),
transequatorial magnetic loops, transequatorial filaments and long
filaments along the boundaries of EBRs (maybe corresponding to the
high-latitude CMEs). The latitudes of former three categories are
higher or lower than active regions, which result in a much wider
distribution of CME latitudes.

\section{Conclusions}

Using the proposed method, we have studied the solar cycle
variation of real CME latitudes during cycle 23 and found three
main features. (1) The latitudinal drift of the high-latitude CMEs
($\theta>60^{\circ}$) was correlated with that of the HCS. (2) The
mid-latitude CMEs ($\theta\in[45^{\circ},60^{\circ}]$) occurred
primarily in 2000, near the middle of 2002 and in 2005,
respectively, forming a prominent three-peak structure. (3) The
latitudinal evolution of low-latitude CMEs ($\theta< 45^{\circ}$)
did not follow the Sp\"{o}rer sunspot law, which suggests that
many CMEs originated outside of active regions. We believe that
such features can provide hints for understanding the origin of
CMEs. At present, Solar Terrestrial Relations Observatory
(STEREO), a very nice mission of NASA and launched successfully
last October, is observing solar corona in 3-D instead of on the
plane of the sky. With it the finding of CME source regions is no
longer a difficult thing and then real CME latitude distribution
can be better determined.

\acknowledgments

This CME catalog is generated and maintained at the CDAW Data
Center by NASA and The Catholic University of America in
cooperation with the Naval Research Laboratory. SOHO is a project
of international cooperation between ESA and NASA. This work is
jointly supported by National Natural Science Foundation of China
(40621003, 40536029, and 40604019), the 973 project under grant
2006CB806304, and the CAS International Partnership Program for
Creative Research Teams.

\clearpage

\begin{figure}
\plotone{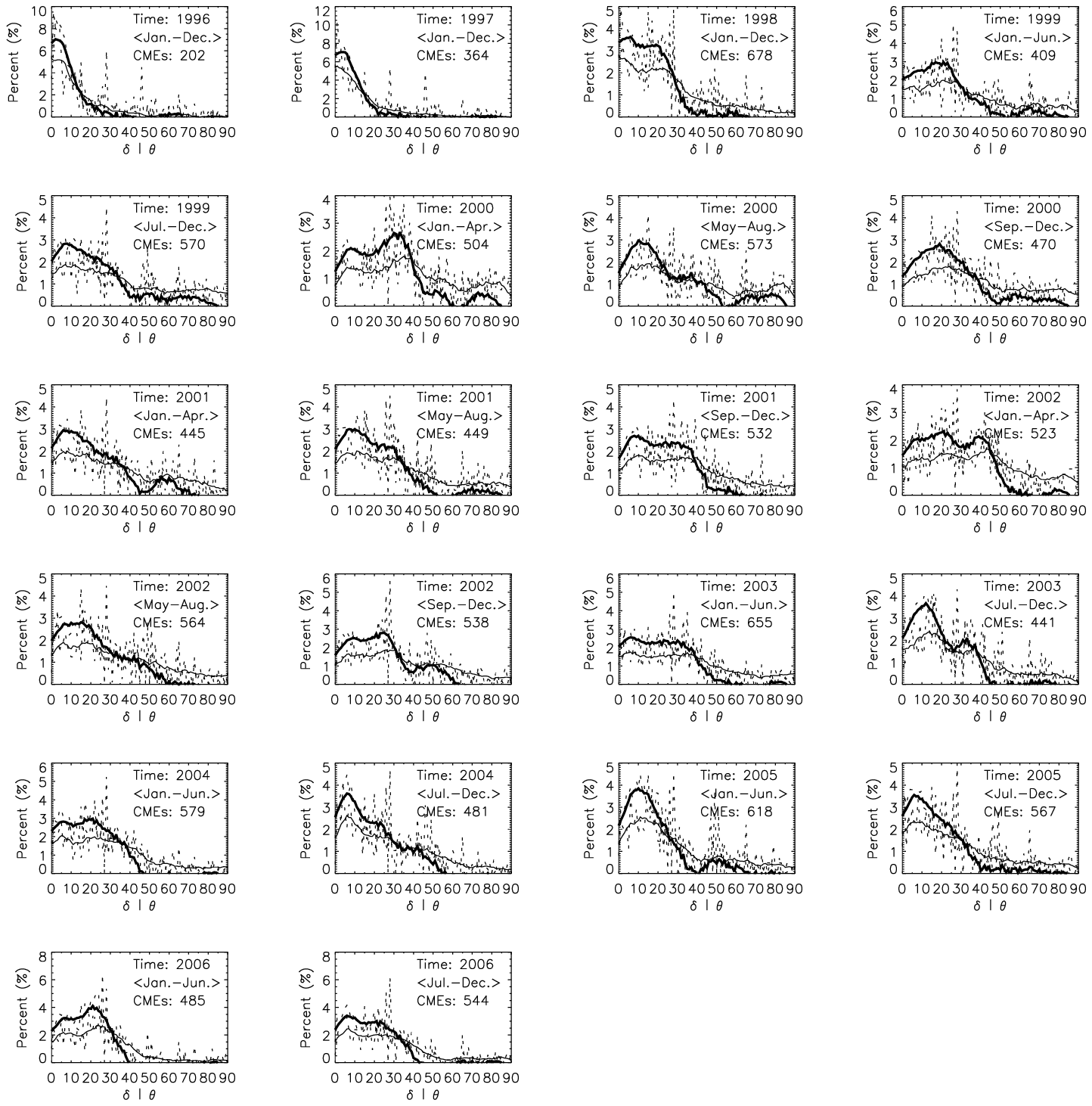} \caption{The distributions of the apparent ($S$,
thin solid line) and real ($R$, thick solid line) CME latitudes
during 22 time intervals which are smoothed with a window of
$10^{\circ}$. The dashed lines around them indicate the unsmoothed
variations. In the top right corner of each plot, the time range
and the CMEs number have been shown. \label{f1}}
\end{figure}

\begin{figure}
\plotone{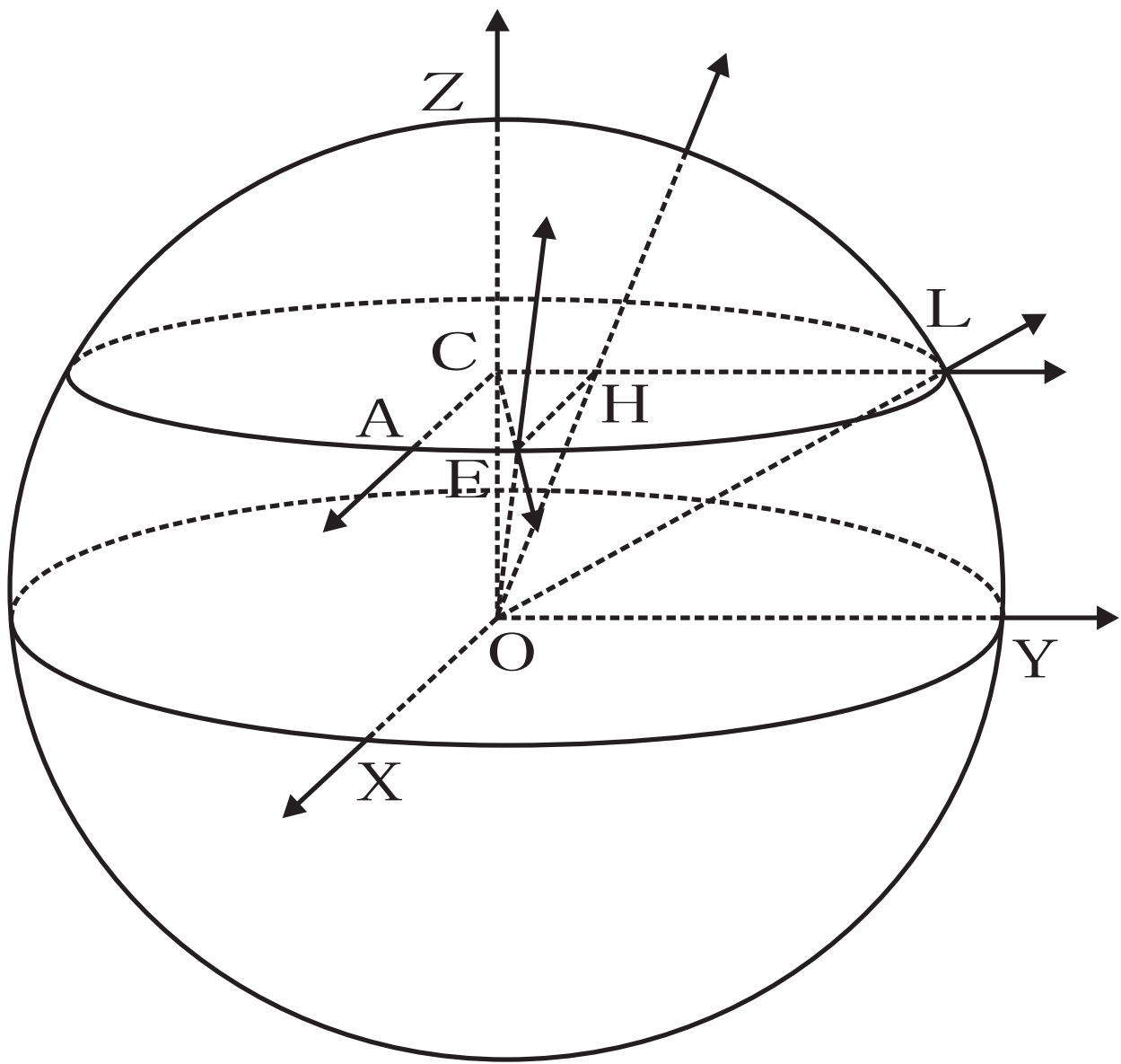} \caption{The spatial sketch of the apparent
latitude ($\delta$, $\angle{HOY}$), the real latitude ($\theta$,
$\angle{LOY}$) and the longitude ($\varphi$, $\angle{ACE}$) of a
CME event initiating from the point $E$. $OX$ is the direction of
the observer, $OE$ is the direction of CME motion, $CA//OX$,
$CL//OY$, and $EH\perp CL$. \label{f1}}
\end{figure}

%\clearpage

\clearpage
\begin{figure}
\plotone{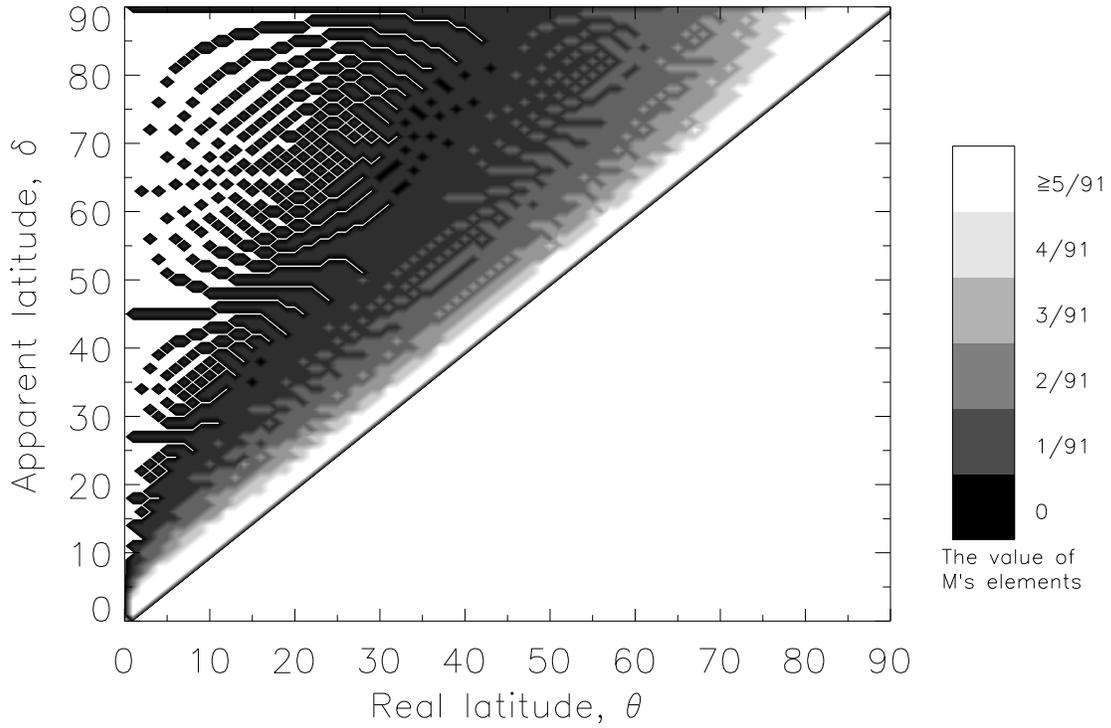} \caption{The contour of the transition matrix $M$
defined in section 3. The distributions of the apparent ($S$) and
real ($R$) CME latitudes are related by the equation $S=M\times
R$. Bright white in the upper left and down right corners
indicates the value of zero only for being pleasing to the eye.
The bin widths of $\theta$ and $\delta$ are both $1^{\circ}$.
\label{f1}}
\end{figure}

\clearpage
\begin{figure}
\plotone{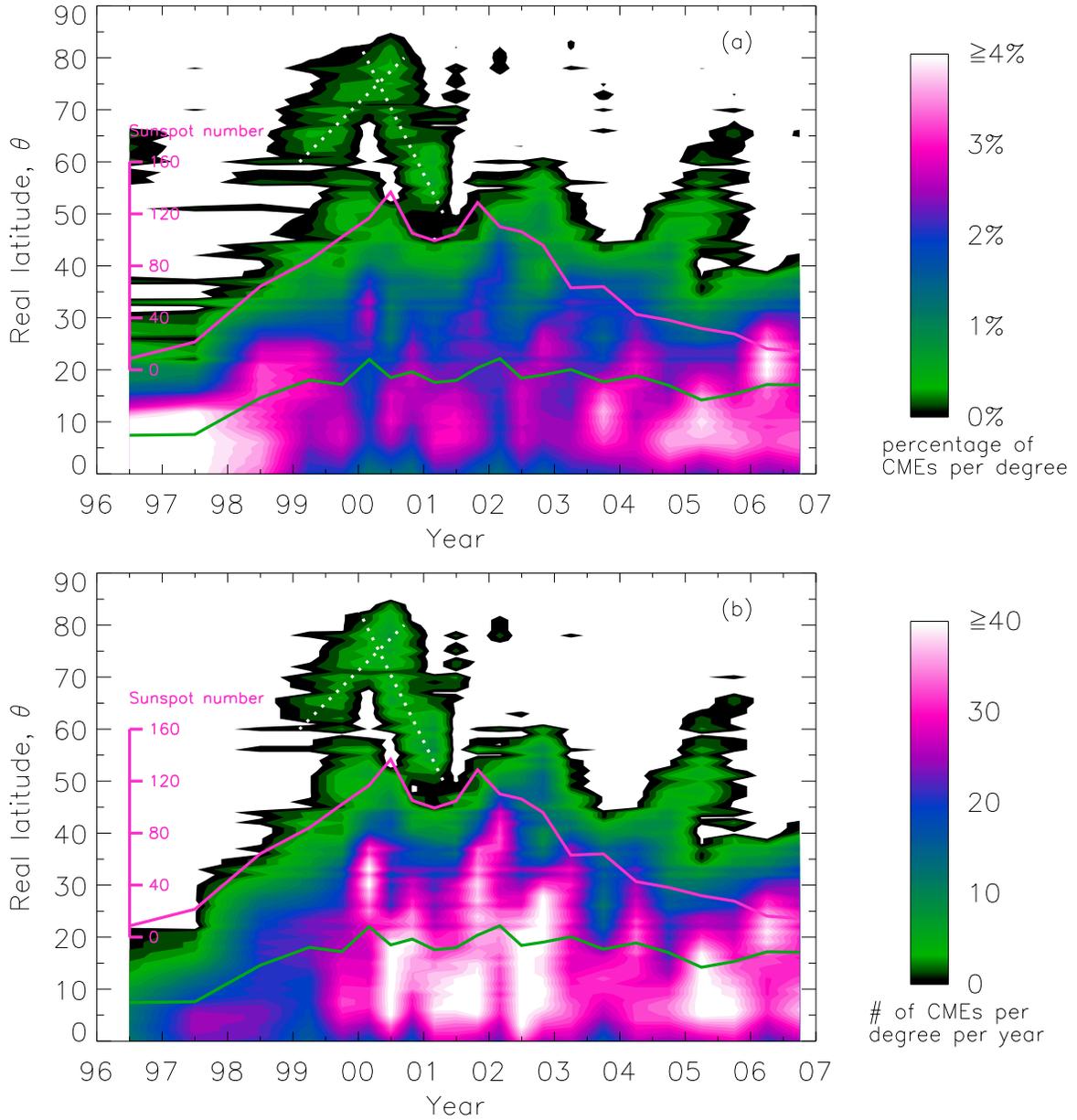} \caption{The solar cycle variations of the real
CME latitude $\theta$ (a) and the CME occurrence rate $\eta$ (b).
The white dotted lines show the poleward and equatorward motions
of the high-latitude CMEs. The pink solid line indicates the
monthly averaged sunspot number during each period of time. The
green solid line indicates the average latitude of CMEs in the
range $\theta<45^{\circ}$. Bright white in the upper regions
indicates the value of zero. \label{f1}}
\end{figure}

\end{document}